\documentclass{ws-ijmpa}
\usepackage[super,compress]{cite}
\usepackage{graphicx}
\begin{document}
\markboth{Kerson Huang, Chi Xiong and Xiaofei Zhao}
{Scalar-field theory of dark matter}

%
\catchline{}{}{}{}{}
%

\title{Scalar-field theory of dark matter
}

\author{Kerson Huang}

\address{Physics Department, Massachusetts Institute of Technology, \\Cambridge, MA 02139 USA\\
kerson@mit.edu}

\author{Chi Xiong}

\address{Institute of Advanced Studies, Nanyang Technological University, \\Singapore 639673\\xiongchi@ntu.edu.sg}

\author{Xiaofei Zhao}

\address{Department of Mathematics, National University of Singapore, \\Singapore 119076\\a0068978@nus.edu.sg }

\maketitle


\begin{abstract}
We develop a theory of dark matter based on a previously proposed picture, in
which a complex vacuum scalar field makes the universe a superfluid, with the
energy density of the superfluid giving rise to dark energy, and variations
from vacuum density giving rise to dark matter. We formulate a nonlinear
Klein-Gordon equation to describe the superfluid, treating galaxies as
external sources. We study the response of the superfluid to the galaxies, in
particular, the emergence of the dark-matter galactic halo, contortions during
galaxy collisions, and the creation of vortices due to galactic rotation.

\end{abstract}

\ccode{PACS numbers: 98.80.-k, 95.35.+d, 11.10.Lm, 03.75.Kk}


\section{Introduction and summary}

The standard model of particle theory postulates the existence of at least one
complex vacuum scalar field (the Higgs field), in order to generate certain
masses in a gauge-invariant manner, mainly in the weak sector, through
spontaneous symmetry breaking. Direct experimental support of this field comes
from the discovery of the associated field quantum, the Higgs boson \cite{Higgs1, Higgs2}.
Extension of the standard model to supersymmetric or grand unified theories
will bring in still more complex vacuum fields. We may therefore suppose that
the universe is permeated with complex scalar fields, whose manifestation in
the large has not been the preoccupation of particle theory, but is the main
focus of this paper.

As we learn from condensed matter physics, a complex scalar field on the
macroscopic scale gives rise to superfluidity. The field serves as an order
parameter that expresses quantum phase coherence over macroscopic distances,
and the superfluid velocity corresponds to the gradient of the phase of the
complex field. From this point of view, we must view the universe as
superfluid. It has been proposed \cite{HLT1, HLT2} that this cosmic superfluidity offers
explanations to both ``dark energy" and ``dark matter":

\begin{itemize}
\item Dark energy is the energy density of the cosmic superfluid, which drives
an accelerated expansion of the universe;

\item Dark matter is the manifestation of local density variations of the
superfluid, which are detectable through gravitational lensing.
\end{itemize}

\noindent In this paper, we formulate and apply a phenomenological theory of
dark matter based on this picture.

In Ref.\cite{HLT1}, the dark-energy aspect is explored in a model using a
self-interacting scalar field coupled to gravity through Einstein's equation
with Robertson-Walker metric, in the era immediately following the big bang.
The Robertson-Walker metric sets the only length scale in the problem, and the
field potential must change with this scale through quantum field-theoretic
renormalization. For mathematical consistency, one requires that the potential
be asymptotically free (i.e., vanish in the small-distance limit,) and this
determines it to be the Halpern-Huang potential \cite{Halpern-Huang} derived from
renormalization-group studies. It is a non-polynomial potential with
exponential behavior for large fields, a feature leading to the generation of
an effective cosmological constant that decrease in time like a power, and
this gives dark energy (accelerated expansion of the universe), while avoiding
the usual ``fine-tuning" problem.

The same model gives a new scenario of the inflation era \cite{HLT2}. After the big
bang, matter (modeled as a perfect fluid coupled to the scalar field) was
created in the quantum turbulence of the cosmic superfluid, in the form of a
vortex tangle. It grows and decays in about 10$^{-23}$s, and is described
phenomenologically by Vinen's equation originally used in liquid helium. The
inflation era corresponds to the lifetime of quantum turbulence, during which
the radius of the universe increases by a factor $10^{27}$, and all the matter
needed for subsequently nucleosynthesis was created. At the end of the
inflation era, the usefulness of this model is ended, for density variations
in the universe became important. The model passes control to the standard hot
big bang theory, with an important condition: the universe remains a
superfluid, and all astrophysical processes take place in this superfluid.

The foregoing describes how the cosmic superfluid comes into being, and its
role during the inflation era. Now we have to describe it post inflation, in
which conditions have become somewhat different.

The big bang era was governed by the Planck scale of 10$^{18}$GeV, which was
built into Einstein's equation through the Robertson-Walker metric. With the
emergence of matter comes the nucleon mass scale of 1 GeV. This scale
ultimately arises in scale-invariant QCD through spontaneous breaking of
chiral symmetry --- an effect referred to as ``dimensional transmutation". (The
observable mass of the universe comes mainly from protons and neutrons, and
has little to do with the Higgs.) Since the matter length scale is much
smaller than the Planck scale, it should prevail in the dynamics of stars and
galaxies. The decoupling between the nuclear scale and the Planck scale is
illustrated in Ref.\cite{HLT2}. In the present work, which is largely concerned with
the present universe, we simply ignore the influence of the Planck scale, and
use a simple $\phi^{4}$ theory with constant phenomenological coupling constants.

In principle, one should use an updated Halpern-Huang potential, with needed
renormalization to bridge the Planck era and the present. Everything that
happens between then and now --- the expansion of the universe, emergence of
the matter scale, interactions between matter and the scalar field --- will
contribute to this ``running" of the potential, with the length scale expanding
by 18 orders of magnitude. We bypass this problem here by adopting the
phenomenological $\phi^{4}$ potential.

The order parameter originates in Landau's \cite{Landau} theory of phase transitions, in
which a high-temperature phase with less ``order" (i.e., higher degree of
symmetry) changes into a low-temperature phase with more ``order" (less
symmetry), through what is now called spontaneous symmetry breaking. The order
parameter is something that gives a measure of ``order", and it vanishes above
the critical temperature. The generic example is the magnetic moment density
in a paramagnetic phase transition, which can be represented by a scalar field
$\phi$ with a symmetric field potential $V(\phi)$, which has only one minimum
with $\phi=0$ at high temperatures, but exhibits two minima at nonzero values
$\phi=\pm\phi_{0}$ below the critical temperature. This is a phenomenological
way to describe spontaneous symmetry breaking. In the Ginsburg-Landau theory
of superconductivity \cite{Ginsburg-Landau}, the order parameter is a complex scalar field
coupled to the electromagnetic field, and it successfully describes the
Meissner effect, in which the photon acquires mass inside a superconductor.
(This is the first instance of the so-called ``Higgs mechanism".) We now know
that this order parameter corresponds to the complex wave function of Cooper
pairs in the microscopic BCS theory \cite{BCS}, and the Ginsburg-Landau theory has
been derived \cite{Gorkov} from BCS. However, it continues to serve as an indispensable
practical tool.

The Landau approach is also important for understanding superfluidity in
liquid helium and cold-trapped atoms \cite{cold-atoms}. The order parameter in this case is
the complex wave function of the Bose-Einstein condensate, and it obeys the
Gross-Pitaevskii equation, an instance of the NLSE (nonlinear Schr\"{o}dinger
equation) first used in nonlinear optics \cite{Kelley}.

In our investigation here, we consider a generic complex scalar field in
curved space-time, which may correspond to some vacuum field in particle
theories, such as supersymmetry or grand unified theories. We assume that the
field obey the nonlinear Klein-Gordon equation (NLKG) in curved space-time. We
shall discuss the definition of superfluid density and superfluid velocity in
a relativistic setting. Matter will be treated as a perfect fluid with given
density current, interacting with the scalar field via current-current
interaction. In this work we have not considered the dynamics of matter itself.

The NLKG describes a pure superfluid at absolute zero temperature. At higher
temperatures, there will appear a normal fluid, consisting of phonons, field
quanta, as well as particles coupled to it, existing with the superfluid in
thermal equilibrium. The so-called WIMPs (weakly interacting massive
particles) searched for in particle-physics setups \cite{PICASSO} would be a component
of our normal fluid, if they exist, and this would explain how they came to
permeate all space. However, we have not taken the normal fluid into account
in this paper.

The most persuasive evidence for dark matter is the detection of galactic
halos via gravitational lensing. In our picture, the dark-matter halo is made
up of extra superfluid drawn around a galaxy, due to attractions between
galaxy and superfluid. The halo can follow the motion of a galaxy like a
``soliton". It can follow the rotation of a galaxy by developing quantized
vortices. When two galaxies collide, their halos will merge and undergo
contortions. All these phenomena are part of the superfluid hydrodynamics that
follows from the NLKG. Observed dark matter formations like the so-called
``bullet cluster" \cite{BulletCluster} and ``train wreck" \cite{TrainWreck} are, in our picture,
manifestations of superfluid hydrodynamics.

The main topics investigated in this paper are
\begin{itemize}
\item fitting model parameters to cosmological data,

\item dark matter in galactic halos,

\item quantized superfluid vortices,

\item numerical simulations of dark-matter halos, galaxy collisions, vortices.
\end{itemize}

\section{Nonlinear Klein-Gordon equation (NLKG)}

We use units in which $\hbar=c=1$, and a metric corresponding to $(-1,1,1,1)$
in flat space-time. The vacuum complex scalar field is denoted by
\begin{equation}
\phi\left(  x\right)  =F\left(  x\right)  e^{i\sigma\left(  x\right)  }
\end{equation}
The classical Lagrangian density in the absence of galactic matter is
\begin{equation}
\mathcal{L}=-g^{\mu\nu}\partial_{\mu}\phi^{\ast}\partial_{\nu}\phi-V
\end{equation}
where $g^{\mu\nu}$ is the metric tensor, and $V\left(  \phi^{\ast}\phi\right)
$ is the self-interaction potential. The action is
\begin{equation} \label{phiaction}
S=-\int dt\,d^{3}x\sqrt{-g}\left(  g^{\mu\nu}\partial_{\mu}\phi^{\ast}
\partial_{\nu}\phi+V\right)
\end{equation}
where $g=$ det$\left(  g_{\mu\nu}\right)  $. This leads to the equation of
motion
\begin{equation}
\frac{1}{\sqrt{-g}}\partial_{\mu}\left(  \sqrt{-g}g^{\mu\nu}\partial_{\nu}
\phi\right)  -\frac{\partial V}{\partial\phi^{\ast}}=0
\end{equation}
We use a phenomenological $\phi^{4}$ potential, which is the simplest way to
maintain a nonzero vacuum field $F_{0}$:
\begin{equation} \label{scalarpotential}
V=\frac{\lambda}{2}\left(  \left\vert \phi\right\vert ^{2}-F_{0}^{2}\right)
^{2}+V_{0}
\end{equation}
where $V_{0}$ is the vacuum energy density, the effective cosmological
constant that gives rise to dark energy. It is easy to see that the action (\ref{phiaction}) has a global U(1) symmetry since the potential $V$ is invariant under the transformation 
$\phi \rightarrow \phi e^{i \alpha}, ~\phi^* \rightarrow \phi^* e^{-i \alpha}$.

In flat space-time, in the absence of galactic matter, the equation of motion
is the nonlinear Klein-Gordon equation (NLKG)
\begin{equation}
\left(  \nabla^{2}-\frac{\partial^{2}}{\partial t^{2}}\right)  \phi
-\lambda\left(  \left\vert \phi\right\vert ^{2}-F_{0}^{2}\right)  \phi=0
\end{equation}
with a conserved current density
\begin{equation} \label{current}
j^{\mu}=\frac{1}{2i}\left(  \phi^{\ast}\partial^{\mu}\phi-\phi\partial^{\mu
}\phi^{\ast}\right)  =F^{2}\partial^{\mu}\sigma
\end{equation}
In the presence of a galaxy (a generic term that includes a star), we add to
the Lagrangian density an interaction term $\mathcal{L}_{\text{int}}$, which
represents the non-gravitational interaction between galaxy and scalar field.
(Gravity will be included in the next section).
The galaxy is introduced as an external source, with a given energy current
density $J^{\mu}$. Its interaction with the scalar field is described
phenomenologically through a current-current interaction, as dictated by
Lorentz covariance:
\begin{equation} \label{Jmujmu}
\mathcal{L}_{\text{int}}=-\eta J^{\mu}j_{\mu}
\end{equation}
where $\eta$ is a coupling constant, and $J^{\mu}$ is a four-vector:
\begin{equation}
J^{\mu}=\left(  \rho,\mathbf{J}\right)
\end{equation}
where the given function $\rho\left(  x\right)  $ describes the density
profile of the galaxy. 
Note that from (\ref{current}) and (\ref{Jmujmu}) we have $\mathcal{L}_{\text{int}} = -\eta F^2 J^{\mu} \partial_{\mu} \sigma$ which has, from particle physics point of view, the typical form of a coupling between the Noether current $J^{\mu}$ from the normal matter sector and the Nambu-Goldstone boson $\sigma$, if the U(1) symmetry is spontaneously broken. In fact $\eta F_0^2$ can be parametrized as $f_{\phi}^{-1}$ in analogy with the pion decay constant $f_{\pi}$. In this article we shall treat it phenomenologically as an effective coupling at the macroscopic level, and fit it to galactic data. (See section 8 for the estimate of the range of $\eta$, and section 9 for the discussion on the long-range force induced by the scalar $\phi$ ).

As an example we consider the rotation of a galaxy submerged in the cosmic superfluid. For a galaxy rotating as a rigid body with angular velocity $\mathbf{\Omega}$,
we have
\begin{equation}
\mathbf{J}=\rho\mathbf{\Omega\times r}
\end{equation}
where $\mathbf{r}$ is the distance vector from the center of the galaxy. In
the equation of motion, $\eta$ and $\rho$ occur only in the combination
$\eta\rho.$ We define the coupling constant $\eta$ precisely by normalizing
$\rho$ :
\begin{equation}
\int d^{3}x\,\rho=M_{\text{galaxy}}
\end{equation}
where $M_{\text{galaxy}}$ is the total mass of the galaxy. 

The NLKG now reads
\begin{equation}
\left(  \nabla^{2}-\frac{\partial^{2}}{\partial t^{2}}\right)  \phi
-\lambda\left(  \left\vert \phi\right\vert ^{2}-F_{0}^{2}\right)  \phi-i\eta
J^{\mu}\partial_{\mu}\phi=0
\end{equation}
or
\begin{equation}
\left(  \nabla^{2}-\frac{\partial^{2}}{\partial t^{2}}\right)  \phi
-\lambda\left(  \left\vert \phi\right\vert ^{2}-F_{0}^{2}\right)  \phi
-i\eta\rho\left(  \frac{\partial\phi}{\partial t}+\mathbf{\Omega\times r}
\cdot\nabla\phi\right)  =0
\end{equation}
This equation contains a correlation length $\xi_{0}$ arising from the vacuum
field $F_{0}$:
\begin{equation}
\xi_{0}\equiv\frac{1}{\sqrt{\lambda}F_{0}}
\end{equation}
which we identify with the correlation length in galaxy clustering \cite{Adelberger}
\begin{equation}
\xi_{0}\approx11\text{ Mpc}=3.3\times10^{25}\text{cm} \label{corr00}
\end{equation}
This gives the mass of scalar-field quanta as
\begin{equation}
m_{0}=\xi_{0}^{-1}\approx1.8\times10^{-30}\text{ eV} \label{quanta}
\end{equation}

Another estimate of $\xi_{0}$ can be obtained from the power spectrum of the
CMB (cosmic microwave background). When analyzed in terms of Legendre
polynomials $P_{\ell}\left(  \cos\theta\right)  $, the angular distribution of
thermal fluctuations in the CMB shows a peak at $\ell\approx200$ \cite{CMB}. Now,
the angular pattern $P_{\ell}\left(  \cos\theta\right)  $ divides the circle
into $2\ell$ sectors, each subtending an angle $\Delta\theta\approx\pi/\ell$ ,
and stuff within this angle should be correlated. So the correlation length
would be of the order of $\xi_{0}\sim\pi R/\ell,$where $R\sim10^{28}$ cm is
the distance of last scattering. Putting $\ell=200$, we get $\xi_{0}
\sim10^{26}$ cm, which is not inconsistent with (\ref{corr00}), considering
the qualitative nature of the argument.

The NLKG here is a field equation, and is quite different from the
relativistic N-body equations used in astrophysics. The scalar field here is
the vacuum expectation value of a quantum field, which contains a high-
momentum cutoff $\Lambda$. When $\Lambda$ changes, the coupling $\lambda$
``renormalizes", in order to maintain the identity of the theory. For the
$\phi^{4}$ theory, $\lambda\left(  \Lambda\right)  \propto\ln\Lambda$. We take
$\Lambda=\Lambda_{\text{QCD}}+\Lambda_{\text{cosmo}},$ where $\Lambda
_{\text{QCD }}$refers \ to the nuclear energy scale, and is taken to be
time-independent, while $\Lambda_{\text{cosmo}}$ is the energy scale of the
space-time metric, and decreases with the expanding universe. The latter is
ignored because $\Lambda_{\text{cosmo}}<<\Lambda_{\text{QCD}}$. This is why we
can treat $\lambda$ as a constant.

\section{Gravitational interactions}

We now include gravitational interactions in the Newtonian limit. The metric
tensor is given by
\begin{align}
g_{00}  &  =-\left(  1+2U\right)  \text{ \ \ \ }\left(  U<<1\right)
\nonumber\\
g^{00}  &  =-\frac{1}{1+2U}\\
g_{jk}  &  =\delta_{jk}\nonumber\\
\sqrt{-g}  &  =\sqrt{1+2U}%
\end{align}
where $U$ is the gravitational potential. The equation of motion becomes
\begin{equation}
-\left(  1-2U\right)  \frac{\partial^{2}\phi}{\partial t^{2}}+\nabla^{2}%
\phi+\frac{\partial U}{\partial t}\frac{\partial\phi}{\partial t}+\nabla
U\cdot\nabla\phi-\lambda\left(  \left\vert \phi\right\vert ^{2}-F_{0}%
^{2}\right)  \phi-i\eta\rho\left(  \frac{\partial\phi}{\partial t}%
+\mathbf{\Omega\times r}\cdot\nabla\phi\right)  =0
\end{equation}
The gravitational potential has contributions from the self-gravitation of the
scalar field, and from the galaxy's gravitational attraction:
\begin{align}
U\left(  x\right)   &  =-G\int d^{3}x^{\prime}\frac{\rho_{\text{sc}}\left(
x^{\prime}\right)  +\rho_{\text{galaxy}}\left(  x^{\prime}\right)
}{|x-x^{\prime}|}\nonumber\\
\rho_{\text{sc}}  &  =T^{00}-V_{0}=\left\vert \frac{\partial\phi}{\partial
t}\right\vert ^{2}+\left\vert \nabla\phi\right\vert ^{2}+\frac{\lambda}%
{2}\left(  \left\vert \phi\right\vert ^{2}-F_{0}^{2}\right)  ^{2}\nonumber\\
\rho_{\text{galaxy}}  &  =M_{\text{galaxy}}\rho
\end{align}
where $M_{\text{galaxy}}$ is the galaxy's total mass. With $\hbar=c=1,$ the
gravitational constant $G$ is given by
\begin{equation}
G=\frac{1}{4\pi}\left(  \text{Planck length}\right)  ^{2}=3\times
10^{-66}\text{ cm}^{2}%
\end{equation}

\section{Galactic dark-matter halo}

A galaxy immersed in the cosmic superfluid will become surrounded by a halo of
higher superfluid density than the vacuum. This can be observed through
gravitational lensing, and interpreted by us as dark matter.

To see how the halo is formed, consider the situation in which a hypothetical
galaxy had been formed in the vacuum scalar field long enough for the entire
system to develop into a stationary state. We would have $\phi\varpropto
e^{-i\omega t}$. The eigenfrequency $\omega$ reflects the strength of the
interaction $\eta\rho$, and may be estimated to be
\begin{equation}
\omega\sim\eta\left\langle \rho\right\rangle \label{freq}
\end{equation}
where $\left\langle \rho\right\rangle $ denotes the spatial average of the
galaxy's density, weighted by the squared modulus of the field $\phi^{\ast
}\phi$. This formula can be derived in a variational formulation of the
problem \cite{vortex}. The NLKG would then become
\begin{equation}
\nabla^{2}\phi-\lambda\left(  \left\vert \phi\right\vert ^{2}-F_{0}^{2}
-\frac{\omega^{2}}{\lambda}\right)  \phi-\eta\rho\left(  \omega\phi
+i\mathbf{\Omega\times r}\cdot\nabla\phi\right)  =0 \label{stationary}
\end{equation}
According to this equation, the vacuum field would be changed from $F_{0}$ to
$F_{1}$, with
\begin{equation}
F_{1}^{2}=F_{0}^{2}+\frac{\omega^{2}}{\lambda}%
\end{equation}

The situation described above is of course hypothetical. What actually happens
is that a dark-matter halo will initially develop around the galaxy and
spread, as illustrated schematically in Fig.1. The initial speed of the
spreading is the velocity of light, the character speed in the NLKG, but it
will be slowed or stopped by gravitational interactions. To estimate the
overall average speed, we assume that the halos around galaxies emerged with
their formation, some $10^{10}$ years ago. We take the present radius to be 10
times galactic radius, or 10$^{6}$ ly, judging from gravitational lensing
pictures \cite{BulletCluster}. This gives an average expansion speed less than 10$^{-4}c$.

The parameters $F_{1}$, $\omega$ give rise to new correlation lengths:
\begin{align}
\xi_{1}  &  =\frac{1}{\sqrt{\lambda}F_{1}}\nonumber\\
\xi_{\omega}  &  =\frac{1}{\omega}
\end{align}
and they are related to the vacuum correlation length through
\begin{equation}
\frac{1}{\xi_{1}^{2}}=\frac{1}{\xi_{0}^{2}}+\frac{1}{\xi_{\omega}^{2}}
\label{corr1}
\end{equation}
The region inside the dark-matter halo is governed by $\xi_{1}$, the outside
region is governed by $\xi_{0}$, and there is a transition region between the
two. We may assume that the stationary equation (\ref{stationary})
approximately holds within the halo, but of course it becomes invalid in the
transition region and beyond. When the frequency $\omega$ is sufficiently
large, the halo region will become non-relativistic.

\begin{figure}
[ptb]
\begin{center}
\includegraphics[height=50mm, width=100mm]{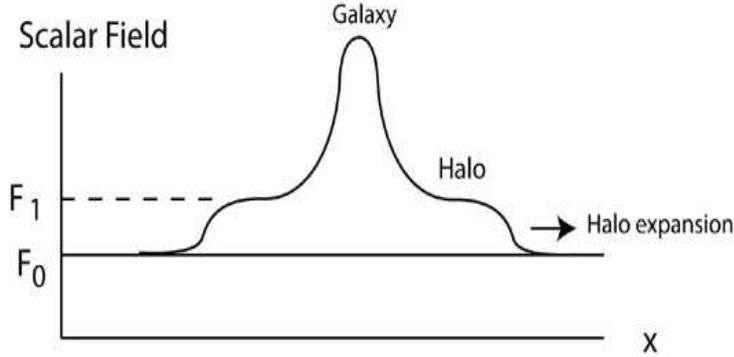}
\caption{Scalar-field profile showing the dark-matter halo around a galaxy.
The vacuum scalar field has modulus $F_{0}$ outside, and $F_{1}>F_{0}$ inside.
This creates a higher energy density in the halo, which can be observed via
gravitational lensing.}%
\end{center}
\end{figure}

For a disk-like galaxy, the halo would emerge as a surface layer on the disk,
and grow in thickness under a force normal to the galactic disk, so as to
lower the energy. Initially the layer would grow at the speed of light, the
characteristic speed of the NLKG, but eventually it would be slowed or stopped
by self gravitation of the layer, and the gravitational attraction from the
galaxy. This could produce the ``squashed beach ball" shape of the halo around
the Milky Way \cite{beachball}. The process is schematically illustrated in Fig.2.

\begin{figure}
[ptb]
\begin{center}
\includegraphics[height=60mm, width=110mm]{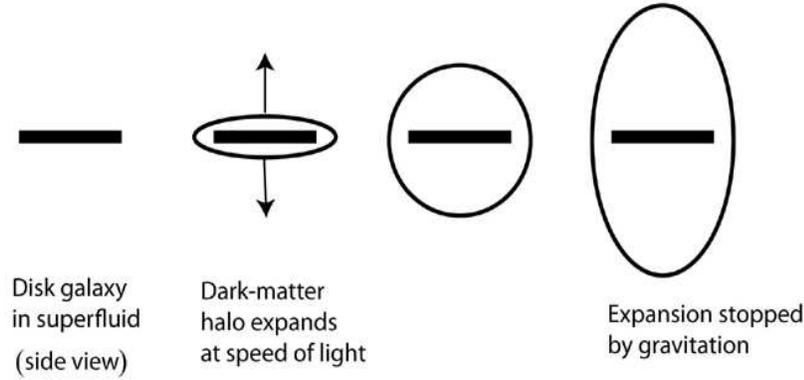}
\caption{Formation of the dark-matter halo around a disk galaxy (side view).
The halo initially grows at the speed of light normal to the galactic disk,
and is eventually stabilized by gravity.}
\end{center}
\end{figure}

\section{Relation between dark energy and dark matter}

The universe roughly consists of 3/4 dark energy and 1/4 dark matter. Since
dark energy and dark matter are manifestations of the same scalar field, that
fact imposes relations that can be tested. In the following we regard dark
matter and halo as the same. Let
\begin{align}
\gamma &  \equiv\frac{\text{Volume of all dark-matter halos}}{\text{Volume of
universe}}\nonumber\\
f  &  \equiv\frac{\rho_{\text{dark matter}}}{\rho_{\text{galaxy}}}
\label{deff}
\end{align}
where $\rho$ denotes mass density. The following relation should hold between
dark energy and dark matter, with neglect of observable matter:
\begin{equation}
\gamma\rho_{\text{dark matter}}+\left(  1-\gamma\right)  \rho_{\text{dark
energy}}=\rho_{0}
\end{equation}
where $\rho_{0}$ is the energy density of the universe. The two terms above
represent roughly 3/4 and 1/4 of the total energy density of the universe,
respectively \cite{Tegmark}. Thus
\begin{align}
\frac{\left(  1-\gamma\right)  \rho_{\text{dark energy}}}{\gamma
\rho_{\text{dark matter}}}  &  =3\nonumber\\
\rho_{\text{dark energy}}  &  =\frac{3\gamma}{1-\gamma}\rho_{\text{dark
matter}}\approx3\gamma\rho_{\text{dark matter}} \label{ratio}
\end{align}
which leads to the relation
\begin{equation}
\gamma=\frac{1}{3}\frac{\rho_{\text{dark energy}}}{\rho_{\text{dark matter}}
}=\frac{1}{3f}\frac{\rho_{\text{dark energy}}}{\rho_{\text{galaxy}}}
\label{gamma}
\end{equation}
Using the values
\begin{align}
\rho_{\text{dark energy}}  &  =V_{0}=5.4\times10^{8}\text{ cm}^{-4}\nonumber\\
\rho_{\text{galaxy}}  &  =1.8\times10^{14}\text{ cm}^{-4} \label{dark}
\end{align}
from \cite{Tegmark} and \cite{Creze} respectively, we get
\begin{equation}
\gamma=\frac{10^{-6}}{f}
\end{equation}
Since each galaxy occupies a volume the size of its halo, we should have
\begin{equation}
\gamma=N_{\text{galaxy}}\left(  \frac{R_{\text{halo}}}{R_{\text{universe}}
}\right)  ^{3}
\end{equation}
where $R_{\text{universe}}$ is the radius of the universe, and
$N_{\text{galaxy}}$ is the number of galaxies. Equating the above to
$f^{-1}10^{-6}$ and then solving for $R_{\text{halo}}$, we get
\begin{equation}
R_{\text{halo}}=\frac{R_{\text{universe}}}{N_{\text{galaxy}}^{1/3}}
\frac{10^{^{-2}}}{f^{1/3}}\sim\frac{10^{5}\text{ly}}{f^{1/3}}
\end{equation}
where we have used the experimental values $R_{\text{universe}}=5\times
10^{28}$cm, $N_{\text{galaxy}}=2\times10^{11}.$ The average radius of a galaxy
happens to be $10^{5}$ly, and the above can be rewritten $R_{\text{halo}}\sim
f^{-1/3}R_{\text{galaxy}}$, or
\begin{equation}
f\sim\left(  \frac{R_{\text{galaxy}}}{R_{\text{halo}}}\right)  ^{3}
\end{equation}
The definition (\ref{deff}) is equivalent to $f=\left(  M_{\text{halo}
}/M_{\text{galaxy}}\right)  \left(  R_{\text{galaxy}}/R_{\text{halo}}\right)
^{3}$, where $M$ denotes total mass. The above thus implies
\begin{equation}
\frac{M_{\text{halo}}}{M_{\text{galaxy}}}\sim1
\end{equation}
From images of gravitational lensing, e.g. the ``bullet cluster" \cite{BulletCluster}, we
estimate $R_{\text{halo}}/R_{\text{galaxy}}\sim10$, (somewhat subjectively,)
and obtain
\begin{align}
f  &  \sim10^{-3}\nonumber\\
\gamma &  \sim10^{-3} \label{fgamma}
\end{align}
This indicated that the halo is a dilute medium with mass density a thousand
times smaller than that of the galaxy.

\section{Superfluid velocity and quantized vortices}

A superfluid can flow pass walls without friction, as long as its velocity is
below a critical velocity, according to the Landau criterion of superfluidity
\cite{Landau}. Similarly, a macroscopic object can move through a superfluid without
friction, as long as its relative velocity is below the critical velocity,
which is usually the phonon velocity associated with long-wavelength
excitations, or Goldstone modes. For our cosmic superfluid, this is the
velocity of light, and thus stars in a galaxy should be able to move
frictionlessly through the dark-matter halo, since the latter is part of the
superfluid. On the other hand, a superfluid can be induced to rotation about
an axis, by creating quantized vortices in the superfluid (See below).

In non-relativistic theories, the superfluid velocity is defined by
$v_{\text{s}}=\left(  \hbar/M\right)  \nabla\sigma$, where $\sigma$ is the
phase of the complex order parameter, and $M$ a mass. However, this is not
bounded by the velocity of light, and clearly needs modification in the
relativistic domain. In flat space-time, we generalize it as follows. We start
with the conserved current $j^{\mu}$ given by (\ref{current}), which satisfies
the continuity equation $\partial_{\mu}j^{\mu}=0$, or
\begin{align}
\frac{\partial n}{\partial t}+\nabla\cdot\mathbf{j}  &  =0\nonumber\\
\mathbf{j}  &  \mathbf{=}F^{2}\mathbf{\nabla}\sigma\nonumber\\
n  &  =-F^{2}\dot{\sigma}
\end{align}
where the field is represent in the form $\phi=Fe^{i\sigma}$, and a dot
denotes time derivative. We interpret $n$ as the superfluid density, and
define the superfluid velocity $\mathbf{v}_{\text{s}}$ through
\begin{equation}
\frac{\partial n}{\partial t}+\nabla\cdot\left(  n\mathbf{v}_{\text{s}
}\right)  =0
\end{equation}
This leads to the definition
\begin{align}
\mathbf{v}_{\text{s}}  &  =\xi_{\text{s}}\mathbf{\nabla}\sigma\nonumber\\
\xi_{\text{s}}  &  \equiv-\frac{1}{\dot{\sigma}}
\end{align}
where $\xi_{\text{s}}$ is a correlation length that is generally space-time
dependent. Now the magnitude of $\mathbf{v}_{\text{s}}$ is bounded by the
velocity of light, as long as $\partial^{\mu}\sigma$ is a time-like four
vector. We recover the nonrelativistic case when $\sigma\left(  \mathbf{r}
,t\right)  =-Mt+\nu\left(  \mathbf{r},t\right) ,$ with $M\rightarrow\infty$.

It should be noted that $\mathbf{v}_{\text{s}}$ has physical meaning as a
velocity field only when $\xi_{\text{s}}$ is constant, or at least slowing varying.

For a manifestly Lorentz-covariant formulation, we introduce a time-like unit
four vector
\begin{equation}
W^{\mu}=\kappa\partial^{\mu}\sigma
\end{equation}
with $W^{\mu}W_{\mu}=-1$. Thus $\kappa$ is given by
\begin{equation}
\kappa=\frac{1}{\sqrt{\dot{\sigma}^{2}-\left\vert \mathbf{\nabla}
\sigma\right\vert ^{2}}}
\end{equation}
and
\begin{align}
\mathbf{W}  &  =\frac{\mathbf{v}_{\text{s}}}{\sqrt{1-\mathbf{v}_{\text{s}}
^{2}}}\nonumber\\
W^{0}  &  =\frac{\mathbf{1}}{\sqrt{1-\mathbf{v}_{\text{s}}^{2}}}
\end{align}

To discuss vorticity, it is more convenient to work with $\partial^{\mu}
\sigma$ without the non-constant factor $\kappa,$ because it is $\partial
^{\mu}\sigma$ that satisfies a quantization condition. Denote the spatial
gradient by
\begin{equation}
\mathbf{u}=\nabla\sigma=\mathbf{v}_{\text{s}}/\xi_{\text{s}}
\end{equation}
The line integral of $\mathbf{u}$ over a closed loop $C$ must be a multiple of
$2\pi$, by continuity of the scalar field:
\begin{equation}
\oint_C d\mathbf{s\cdot u}= \oint_C d\mathbf{s\cdot}\nabla\sigma=2\pi N
\end{equation}

When $N\neq0$, there is a vortex line encircled by $C$. We refer the reader to
Ref.\cite{HLT2} for a discussion of vortex dynamics.

\begin{figure}
[ptb]
\begin{center}
\includegraphics[height=60mm, width=90mm]{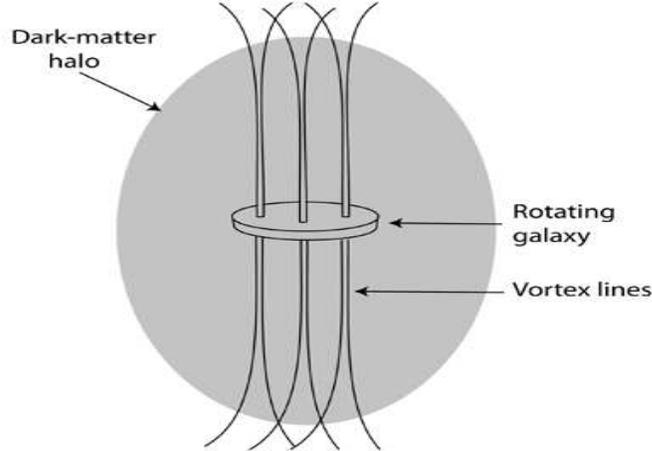}
\caption{Core size of vortex created by a rotating galaxy is governed by
the local correlation length. Inside the galactic it is some 60 times smaller
than that in the vacuum, according to (\ref{Fratio}). Consequently, the size
of the vortex core vastly increases towards the edge of the halo.}%
\end{center}
\end{figure}

For disk-like galaxies rotating about an axis normal to the disk, vortex lines
will be generated normal to the disk, and there is cylindrical symmetry at
least near the surface of the galactic disk. The core size of the vortex line
is proportional to the correlation length, which is $\xi_{1}$ inside the
galactic halo, and increases as we go away from the center of the halo, and
approach the larger vacuum value $\xi_{0}$ outside the halo. The transition
layer between the halo and the vacuum may be regarded as a fuzzy boundary
where a vortex line can ``terminate". This is illustrated in Fig.3. The halo of
a galaxy with one vortex would look like a fat donut with a tiny hole. The
picture might be different for a black hole, for which the strong space-time
curvature may have important effects on vorticity.

We can roughly estimate the critical angular velocity for the creation of one
vortex, using a cylindrical geometry. Consider a circular loop $C$ of radius
$R$, with an infinite vortex line normal to the loop through its center. The
velocity $\mathbf{u}$ is tangential to the circle with constant magnitude
$u_{0}$. We find by explicit evaluation
\begin{equation}
\oint_{C} d\mathbf{s\cdot u}=2\pi Ru_{0}
\end{equation}
Equating this to $2\pi N$ yields $u_{0}=N/R,$ or
\begin{equation}
\frac{v_{\text{s}}}{c}=\frac{\xi_{\text{s}}N}{R}
\end{equation}
The corresponding angular velocity is $\Omega=v_{\text{s}}/R$. Thus, to have
one vortex inside radius $R$, the angular velocity must exceed the critical
value
\begin{equation}
\Omega_{\text{c}}=\frac{c\xi_{\text{s}}}{R^{2}}
\end{equation}
In Table 1, we give estimates of $\Omega_{\text{c}}$ for various stellar
objects, putting $\xi_{\text{s}}\sim R$, i.e., assuming that the correlation
length of the superfluid is greater than the size of the object. The actual
angular velocities $\Omega$ are also cited for reference. As we can see, a
black hole could generate a million vortices, the Milky Way and neutron star
are marginally critical cases, while the sun and earth fall short of criticality.

When a black hole ``drags" the surrounding superfluid into rotation, there
would appear the order of 10$^{6}$ vortex lines, which would form a cage
around the black hole. This may explain the so-called ``non-thermal filaments"
observed in the neighborhood of Sagitarus near the center of the Milky Way \cite{LaRosa}.%
\begin{table}[tbp] \centering
\begin{tabular}
[c]{llll} \toprule
& $R\text{ (cm)}$ & $\Omega_{\text{c}}\text{ (rad/s)}$ & $\Omega\text{
(rad/s)}$\\ \colrule
$\text{Milky Way}$ & $10^{24}$ & $10^{-14}$ & $10^{-15}$\\
$\text{Earth}$ & $10^{9}$ & $10$ & $10^{-4}$\\
$\text{Sun}$ & $10^{11}$ & $10^{-1}$ & $10^{-4}$\\
$\text{Neutron star}$ & $10^{6}$ & $10^{4}$ & $10^{3}$\\
$\text{Black hole}$ & $10^{9}$ & $10$ & $10^{6}$ \\ \botrule
\end{tabular}
\caption{Critical angular velocity for creation of one vortex. Columns correspond to
radius, critical angular velocity, actual angular velocity.}\label{TableKey}
\end{table}

\section{Determination of model parameters}

Assume that the correlation length $\xi_{1}$ in a typical galactic halo is
roughly the radius of the halo, which we take (somewhat subjectively) to be 10
times that of the disk of a typical galaxy:
\begin{equation}
\xi_{1}\sim5\times10^{23}\text{ cm}
\end{equation}
This makes
\begin{equation}
\frac{F_{1}}{F_{0}}=\frac{\xi_{0}}{\xi_{1}}\sim66 \label{Fratio}
\end{equation}
Since $F_{1}>>F_{0}$, we have $F_{1}^{2}=F_{0}^{2}+\omega^{2}/\lambda
^{2}\approx\omega^{2}/\lambda$, or
\begin{equation}
\omega\approx1/\xi_{1}
\end{equation}

Now we calculate $\omega$ using (\ref{freq}) in the halo, taking $F$ to be
constant, and get
\begin{align}
\omega &  \sim\eta\left\langle \rho\right\rangle =\frac{\eta\int_{\text{halo}
}d^{3}x\rho}{\int_{\text{halo}}d^{3}x}=\frac{\eta M_{\text{galaxy}}
}{V_{\text{halo}}}\\
&  \sim\frac{\eta M_{\text{halo}}}{V_{\text{halo}}}=\eta\rho_{\text{dark
matter}}
\end{align}
where $V_{\text{halo}}$ is the volume of the halo, and we have identified the
halo with dark matter. Using $\omega\approx1/\xi_{1}$ and solving for $\eta$,
we obtain
\begin{equation}  \label{etaFermi}
\eta\sim\frac{1}{\xi_{1}\rho_{\text{dark matter}}}=10^{-35}\text{ cm}
^{3}=10^{4}\text{ fermi}^{3}%
\end{equation}
where fermi $=10^{-13}$ cm is the characteristic length of nuclear interactions. (\ref{etaFermi}) is an interesting and profound result, although it is obtained through a crude estimation. Given the dark matter density that we used in Section V, we see a subatomic scale, $\eta^{1/3} \approx 20 ~\text{fermi} $, emerging from cosmological scales. The value of $\eta^{1/3}$ may vary in different models. For example, if WIMPs are taken into consideration as a normal-fluid component, one may have weakly interacting particles with masses from $10^2$ to $10^3$ GeV which correspond to $\eta^{1/3} \approx 10^{-2} \sim 10^{-3} ~\text{fermi} $. In practice the value of $\eta^{1/3}$ should be constrained by experimental observations (see the next section for discussions on the long-ranged force induced by the scalar field compared with gravity).

We now estimate the vacuum field $F_{0}$. In out model, dark energy comes from
$V_{0}$, and dark matter comes from deviations from the vacuum field:
\begin{equation}
\frac{\rho_{\text{dark energy}}}{\rho_{\text{dark matter}}}=\frac{V_{0}}
{\frac{\lambda}{2}\left(  F_{1}^{4}-F_{0}^{4}\right)  }=3\gamma
\end{equation}
where we have used (\ref{gamma}). Since $F_{1}>>F_{0}$, we have
\begin{equation}
\frac{\lambda F_{1}^{4}}{2V_{0}}=\frac{1}{3\gamma}
\end{equation}
With (\ref{dark}) and (\ref{gamma}), we get
\begin{equation}
\lambda F_{1}^{4}=\frac{2V_{0}}{3\gamma}=3.6\times10^{11}\text{cm}^{-4}
\end{equation}
We can find $F_{1}$ and $\lambda$ separately by noting
\begin{equation}
F_{1}^{2}=\frac{\lambda F_{1}^{4}}{\lambda F_{1}^{2}}=\left(  \lambda
F_{1}^{4}\right)  \xi_{1}^{2}%
\end{equation}
The results are
\begin{align}
F_{1}  &  \sim10^{29}\text{ cm}^{-1}\nonumber\\
F_{0}  &  \sim1.5\times10^{27}\text{ cm}^{-1}\nonumber\\
\lambda &  \sim3.6\times10^{-105} \label{values}
\end{align}

\section{Long-ranged force induced by the scalar field}

The small mass (\ref{quanta}) of the field particle means that the scalar
field would induce a long-ranged force in matter. Consider two static test
sources
\begin{equation}
\rho_{i}\left(  \mathbf{r}\right)  =m_{i}\delta^{3}\left(  \mathbf{r}
-\mathbf{r}_{i}\right)  \text{ \ \ \ \ }(i=1,2)
\end{equation}
The field due to one test source, in the absence of the other, is of the form
$\phi\left(  \mathbf{r},t\right)  =e^{-i\omega_{0}t}F\left(  \mathbf{r}
\right)  ,$ where $\omega_{0}\sim\xi_{0}^{-1}$. The equation for $F$ reads
\begin{equation}
\nabla^{2}F+\left[  \omega_{0}^{2}-\lambda\left(  F^{2}-F_{0}^{2}\right)
\right]  F=\eta\omega_{0}F\rho
\end{equation}
We shall neglect the term in brackets, and put $F=F_{0}$ on the right side.
The field produced by source 1 is then given by
\begin{align}
\nabla^{2}F_{1}\left(  \mathbf{r}\right)   &  =\eta\omega_{0}F_{0}m_{1}%
\delta\left(  \mathbf{r}-\mathbf{r}_{1}\right) \nonumber\\
F_{1}\left(  \mathbf{r}\right)   &  =-\frac{\eta\omega_{0}F_{0}m_{1}}{4\pi
}\frac{1}{\left\vert \mathbf{r}-\mathbf{r}_{1}\right\vert }
\end{align}
The energy of source 2 in this field is given by
\begin{align}
U_{21}  &  =-\eta\int d^{3}x\rho_{2}\text{Im}\left(  \phi_{1}^{\ast}\dot{\phi
}_{1}\right)  =\eta\omega_{0}m_{2}F_{1}^{2}\left(  \mathbf{r}_{2}\right)
\nonumber\\
&  =\frac{\eta^{3}\omega_{0}^{3}F_{0}^{2}m_{1}^{2}m_{2}}{\left(  4\pi\right)
^{2}}\frac{1}{\left\vert \mathbf{r}_{1}-\mathbf{r}_{2}\right\vert ^{2}}
\end{align}
The total potential energy is $U_{\text{scalar}}=U_{12}+U_{21}:$
\begin{align}
U_{\text{scalar}}  &  =\frac{Km_{2}m_{1}}{\left\vert \mathbf{r}_{1}
-\mathbf{r}_{2}\right\vert ^{2}}\nonumber\\
K  &  =\left(  4\pi\right)  ^{-2}\eta^{3}\omega_{0}^{3}F_{0}^{2}\left(
m_{1}+m_{2}\right)
\end{align}
This is to be compared with the gravitational potential energy
\begin{equation}
U_{\text{grav}}=-\frac{Gm_{1}m_{2}}{\left\vert \mathbf{r}_{1}-\mathbf{r}
_{2}\right\vert }
\end{equation}
which dominates at large distances. For $m_{1}=m_{2}=M_{\text{galaxy}}$ and $\eta \sim ~ (20 ~\text{fermi})^3 $ (see eqt. (\ref{etaFermi})),  the two potentials have equal magnitude at distance
\begin{equation}
\left\vert \mathbf{r}_{1}-\mathbf{r}_{2}\right\vert_{eq} =\frac{\left\vert
K\right\vert }{G}\sim10^{19}\text{ cm}
\end{equation}
which is 1\% the thickness of the Milky Way disk. This distance, however, is very sensitive to the value of $\eta$ and can vary enormously in different models. As we mentioned in the previous section, if WIMPs are taken into account, one may have $\eta^{1/3} \sim 10^{-3} ~\text{fermi} $, which reduces the above distance $\left\vert \mathbf{r}_{1}-\mathbf{r}_{2}\right\vert_{eq}$ to $10^{-16} \text{cm}$ or below. This seems to be consistent with the short-distance character of the WIMPs' non-gravitational interactions. Therefore at cosmological scales we may ignore the induced potential in the present work. However, it may have an effect in the distribution of galaxies, and may contribute to the formation of galaxies and galaxy clusters at the early universe.

\section{Goldstone modes and vacuum stability}

Being a gravitating medium, the cosmic superfluid could be unstable against
gravitational clustering. This would be manifested as the spontaneous
emergence of patches of dark matter, analogous to the Jeans instability of a
gravitating fluid \cite{Jeans}. To investigate the stability of the vacuum, we
perturbed it by exciting the Goldstone modes. The NLKG in the absence of
matter, but with gravity in the Newtonian limit, is given by
\begin{equation}
\nabla^{2}\phi-\left(  1-2U\right)  \frac{\partial^{2}\phi}{\partial t^{2}
}+\frac{\partial U}{\partial t}\frac{\partial\phi}{\partial t}+\nabla
U\cdot\nabla\phi-\lambda\left(  \left\vert \phi\right\vert ^{2}-F_{0}
^{2}\right)  \phi=0
\end{equation}
where $U$ is the gravitational potential satisfying the Poisson equation
\begin{equation}
\nabla^{2}U=4\pi G\left[  \left\vert \dot{\phi}\right\vert ^{2}+\left\vert
\nabla\phi\right\vert ^{2}+\frac{\lambda}{2}\left(  \left\vert \phi\right\vert
^{2}-F_{0}^{2}\right)  ^{2}\right]
\end{equation}
The ground state corresponds to $\phi_{0}=F_{0}.$ To investigate the Goldstone
modes, we put
\begin{equation}
\phi=F_{0}+f
\end{equation}
where $f$ is a small perturbation. When gravity is ignored, one can solve the
linearized NLKG via the Bogoliubov transformation
\begin{equation}
f=\alpha e^{-i(\omega t-\mathbf{k\cdot x})}+\beta^{\ast}e^{i(\omega
t-\mathbf{k\cdot x})} \label{Bogo}
\end{equation}
In this case, one obtains the dispersion relation $\omega=k$, as dictated by
Lorentz covariance.

With inclusion of gravity, linearization in $f$ \ leads to
\begin{equation}
\nabla^{2}f-\left(  1-2U\right)  \frac{\partial^{2}f}{\partial t^{2}}
+\frac{\partial U}{\partial t}\frac{\partial f}{\partial t}+\nabla
U\cdot\nabla f-\xi_{0}^{-2}\left(  f^{\ast}+f\right)  =0
\end{equation}
This is a nonlinear equation, because $U$ depends on $f.$ To solve it, we
resort to an approximation through the replacements
\begin{align}
\frac{\partial U}{\partial t}\frac{\partial f}{\partial t}  &  \rightarrow
-i\gamma\frac{\partial f}{\partial t}\nonumber\\
\nabla U\cdot\nabla f  &  \rightarrow-i\mathbf{q\cdot}\nabla f
\end{align}
where $f$ on the right side is given by (\ref{Bogo}), and $\gamma$,
$\mathbf{q}$ are real constants. We assume $U<<1$ and neglect it, and obtain a
set of homogeneous algebraic equations:
\begin{align}
\left[  k^{2}-\omega^{2}+\xi_{0}^{-2}-\left(  \mathbf{q\cdot k}-\gamma
\omega\right)  \right]  \alpha+\xi_{0}^{-2}\beta &  =0\nonumber\\
\xi_{0}^{-2}\alpha+\left[  k^{2}-\omega^{2}+\xi_{0}^{-2}+\left(
\mathbf{q\cdot k}-\gamma\omega\right)  \right]  \beta &  =0
\end{align}
\newline We average over the directions of $\mathbf{q,}$ and take
$\left\langle \mathbf{q\cdot k}\right\rangle =0,$ $\left\langle \left(
\mathbf{q\cdot k}\right)  ^{2}\right\rangle =\frac{1}{2}q^{2}k^{2}.$ Setting
the determinant to zero, we obtain the dispersion relations
\begin{align}
\omega^{2}  &  =k^{2}+\frac{1}{2}\left(  A\pm\sqrt{2Bk^{2}+A^{2}}\right)
\nonumber\\
A  &  =2\xi_{0}^{-2}+\gamma^{2}\nonumber\\
B  &  =q^{2}+2\gamma^{2} \label{freq1}
\end{align}

\begin{figure}
[ptb]
\begin{center}
\includegraphics[height=70mm, width=80mm]{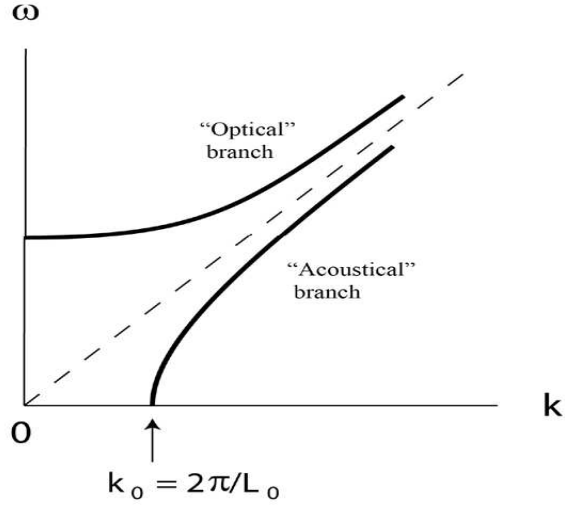}
\caption{The Goldstone modes with $\omega=k$ are split into ``optical" and
``acoustical" branches by gravity. In the acoustical branch, the sound velocity
is pure imaginary in the region $0<k<k_{0}$, and this signifies vacuum
instability, with formation of patches of dark matter of size $L_{0}%
=2\pi/k_{0}.$ In the present universe $L_{0}$ is estimated to be much larger
than the radius of the universe, and $k_{0}\rightarrow0$. Thus, the present
vacuum should be stable against gravitational clustering.}
\end{center}
\end{figure}

The Goldstone modes split into two branches ($\pm$) that are analogous to the
``optical" ($+$) and acoustical ($-$) modes in solids. As depicted in Fig.4,
both revert to the relation $\omega=k$ when $k\rightarrow\infty$. At the low
$k$ end of the spectrum, $\omega$ approaches a finite constant for for the
optical mode, while it goes to zero at $k=k_{0}$ for the acoustical mode, with
the behavior
\begin{align}
&  \omega\approx\sqrt{2k_{0}\left(  k-k_{0}\right)  }\nonumber\\
k_{0}  &  =\sqrt{\frac{1}{2}\left(  B-2A\right)  } \label{k0}
\end{align}
If $k_{0}>0$, this signifies instability, in which the vacuum breaks up into
patches of dark matter with characteristic size
\begin{equation}
L_{0}=\frac{2\pi}{k_{0}}%
\end{equation}

To estimate the instability length $L_{0}$, we use (\ref{k0}) in the form
\begin{equation}
L_{0}=2\pi\left(  \frac{B}{2}-A\right)  ^{-1/2}=2\pi\left(  \frac{q^{2}}
{2}-\frac{2}{\xi_{0}^{2}}\right)  ^{-1/2} \label{L0}
\end{equation}
This is an implicit relation, because $q$ depends on $L_{0}.$ We can estimate
$q$ as follows. Suppose, in spontaneous fluctuation, the vacuum field develops
a value $F_{1}$ in a region of size $L_{0}$. The energy density in this patch
would be $\lambda\left(  F_{1}^{4}-F_{0}^{4}\right)  $, and the mass of the
patch would be
\begin{equation}
M=\lambda\left(  F_{1}^{4}-F_{0}^{4}\right)  L_{0}^{3}
\end{equation}
By dimensional analysis, we estimate that%
\begin{align}
U  &  \sim\frac{GM}{L_{0}}\nonumber\\
q  &  \sim\frac{U}{L_{0}}
\end{align}
Substituting this into (\ref{L0}) yields a quartic equation for $L_{0}$. The
solution in the limit of weak gravity gives
\begin{equation}
\frac{L_{0}}{\xi_{0}}\approx\frac{2}{GF_{0}^{2}\left(  F_{1}^{4}/F_{0}
^{4}-1\right)  }
\end{equation}
Using $G=3\times10^{-66}$ cm$^{2},$ $F_{0}\sim1.5\times10^{27}$ cm$^{-1}$, we
obtain
\begin{equation}
\frac{L_{0}}{\xi_{0}}\approx\frac{3\times10^{12}}{\left(  F_{1}/F_{0}\right)^{4}-1}
\end{equation}
We had obtained $F_{1}/F_{0}\approx66$ in the galactic halo, but that was due
to interaction between galaxy and vacuum field. Here, $F_{1}$ is a property of
the vacuum field, and is determined by balancing the weak gravity against the
strong field potential. We expect that $F_{1}/F_{0}\approx1$, which means
$L_{0}>3\times10^{12}$ $\xi_{0}$. Since this length is overwhelmingly larger
than the radius of the universe, we conclude that the present vacuum is stable
against gravitational clustering. This of course does not preclude vacuum
clustering during the early universe, when conditions were different.

\section{Numerical exploration}

We solve the NLKG with external-source galaxies whose densities have Gaussian
distributions. The galaxies move in prescribed ways as external sources, and
our focus is on the response of the superfluid. Computations on moving
galaxies are carried out in a $512\times512$ two-dimensional spatial grid,
with periodic boundary conditions. Those on vortex lattice creation are done
in a $1024\times1024$ grid. For illustration purposes, model parameters are
chosen arbitrarily.

Fig.5 shows a ``film strip" of the superfluid in the presence of a moving
galaxy, consisting of sequential plots of the scalar-field modulus. Frame 1 is
the initial state, with velocity directed upwards. Frame 2 shows the
development of the dark-matter halo (white area) around the galaxy, and an
outgoing transient cylindrical wave front in the superfluid. In frame 3, the
transient wave continues to expand about the original center, while the halo
moves with the galaxy like a soliton. In frame 4, the transient wave hits the boundaries.

Fig.6 shows the collision between two galaxies, initially made to move towards
each other head-on. They continue on such a course, go through each other, and
finally recede from each other. The frames show the evolution of the field
modulus in contour plots, with the white areas corresponding to the
dark-matter halos. The pictures show the merging of the two halos, the
contortions of the combined halo, and finally the recession of the galaxies
from each other, with their proper halos restored.

Fig.7 shows two galaxies colliding at a nonzero impact parameter. The
superfluid between them is sheared into rotational flow, through the creation
of vortices. In frame 2, two black dots appear, indicating the formation of
vortices, and they continue to develop in the subsequent frames.

Fig.8 shows one moment during the development of a vortex lattice around a
rotating galaxy. Such vortex lattices have been observed in cold-trapped atoms
\cite{Fetter}, and simulated in the NLSE \cite{Tsubota}. To our knowledge, this is the first
simulation in the NLKG. The left panel is a contour plot of the field modulus,
and the right frame is one of the phase of the field. The vortices form four
rings, with the outer most ring located just outside the dark-matter halo (in
white). The vortex counts of the rings are: 19, 12, 11, 9, a total of 51, with
the inner most ring being somewhat irregular because it is still being formed.
In general, vortices are nucleated inside the galaxy, and migrate outwards.
The phase contours show a radial pattern, in which the dark radial spokes
represent ``strings" across which the phase jumps by $2\pi$, and these strings
must be terminated by vortices. For example, the tips of the ``chrysanthemum
petals" coincide with the locations of the vortices on the outermost layer.
Note that vortices in the outermost layer has much larger core size than those
within the halo, because the correlation length is much larger. [See Fig.3 and
Eq.(\ref{Fratio}).]

Fig.9 shows the number of vortices as function the angular velocity of the
galaxy, for a range of the interaction parameter between galaxy and scalar field.%

\begin{figure}
[ptb]
\begin{center}
\includegraphics[height=40mm, width=140mm]{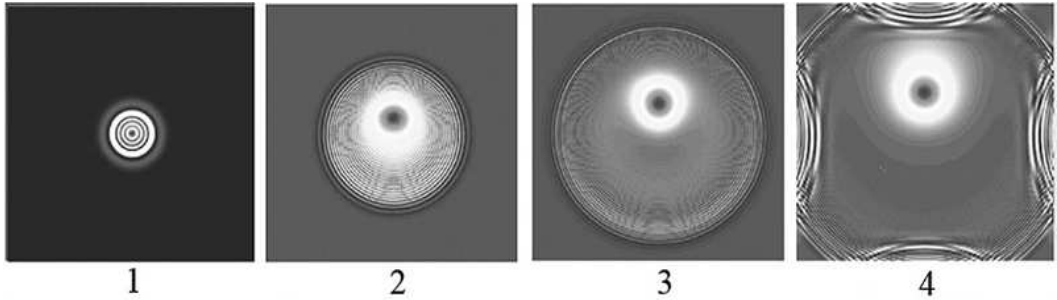}
\caption{Contour plots of the scalar-field modulus in 2D space, showing time
evolution of the superfluid in response to the presence of a moving galaxy.
The white areas correspond to the dark-matter halo. A transient cylindrical
wave front can be seen propagating from the initial position of the galaxy,
hitting the boundaries in the last frame. }
\end{center}
\end{figure}

\begin{figure}
[ptb]
\begin{center}
\includegraphics[height=70mm, width=140mm]{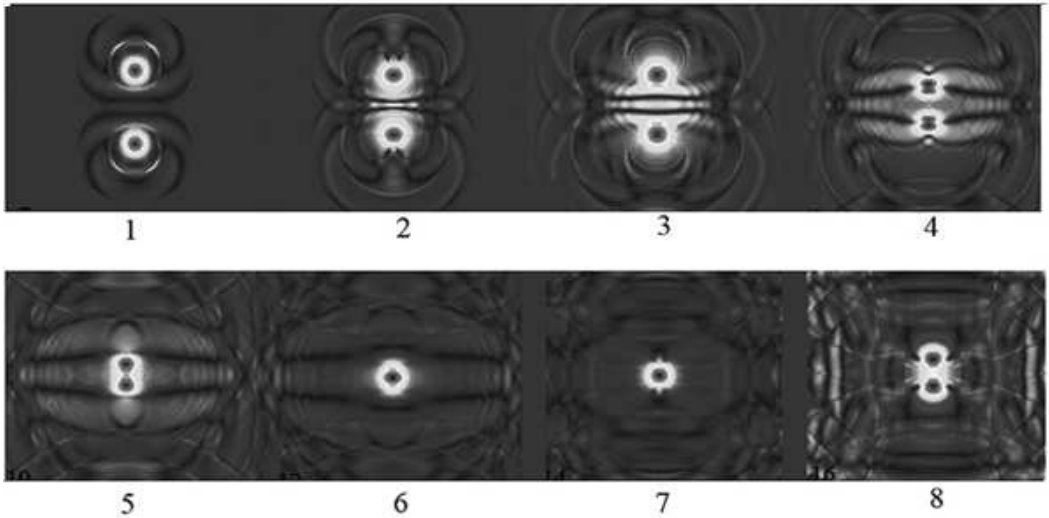}
\caption{Collision between two galaxies approaching each other along the
vertical direction. The dark-matter halos merged, underwent deformations, and
were finally restored when the galaxies recede from each other.}%
\end{center}
\end{figure}

\begin{figure}
[ptb]
\begin{center}
\includegraphics[height=80mm, width=120mm]{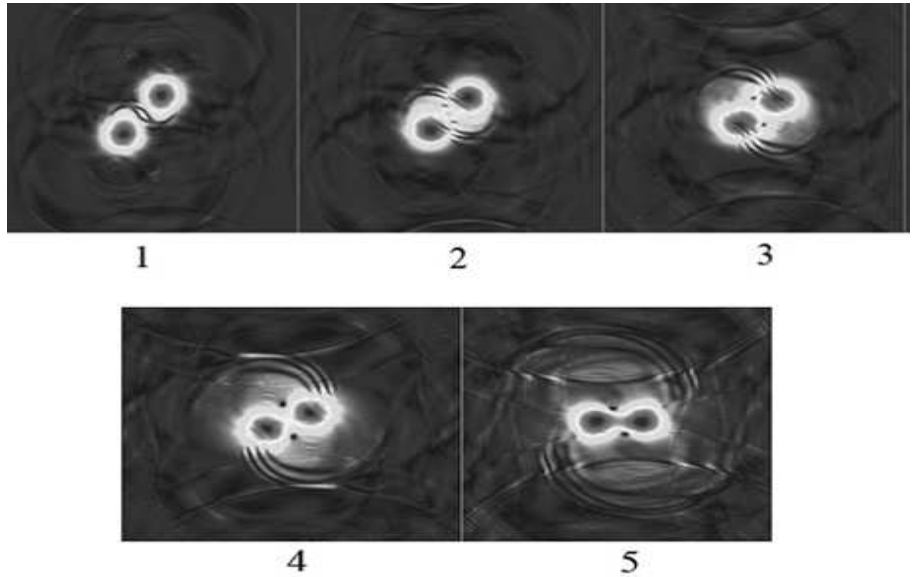}
\caption{Two galaxies passing each other at nonzero impact paramter. The
superfluid between them is sheared into rotation, with creation of vortices
(the black dots).}
\end{center}
\end{figure}

\begin{figure}
[ptb]
\begin{center}
\includegraphics[height=60mm, width=120mm]{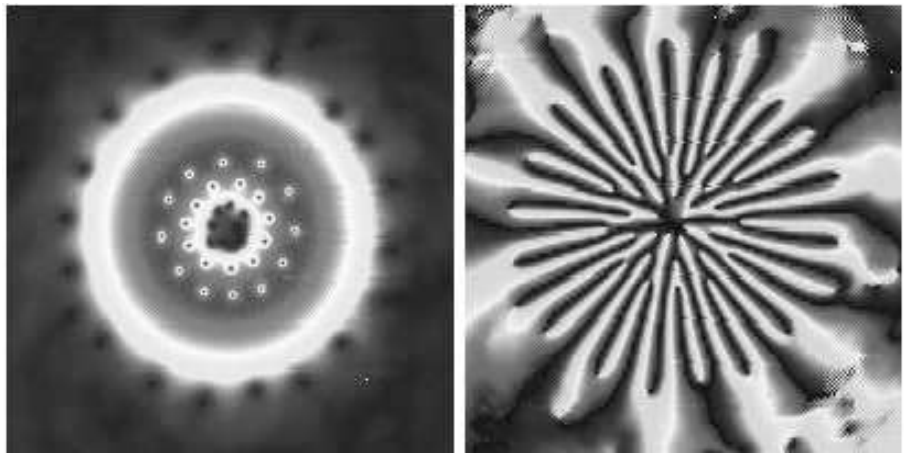}
\caption{A vortex lattice surrounding a rotating galaxy. The left panel is a
contour plot of the field modulus, showing four rings of vortices, the
outermost of which lies beyond the galactic halo, and has a much larger core
size. The right panel is a contour plot of the phase of the field. The radial
spokes are ``strings" across which the phase jumps by $2\pi$. See text for a
fuller description.}
\end{center}
\end{figure}

\begin{figure}
[ptb]
\begin{center}
\includegraphics[height=60mm, width=100mm]{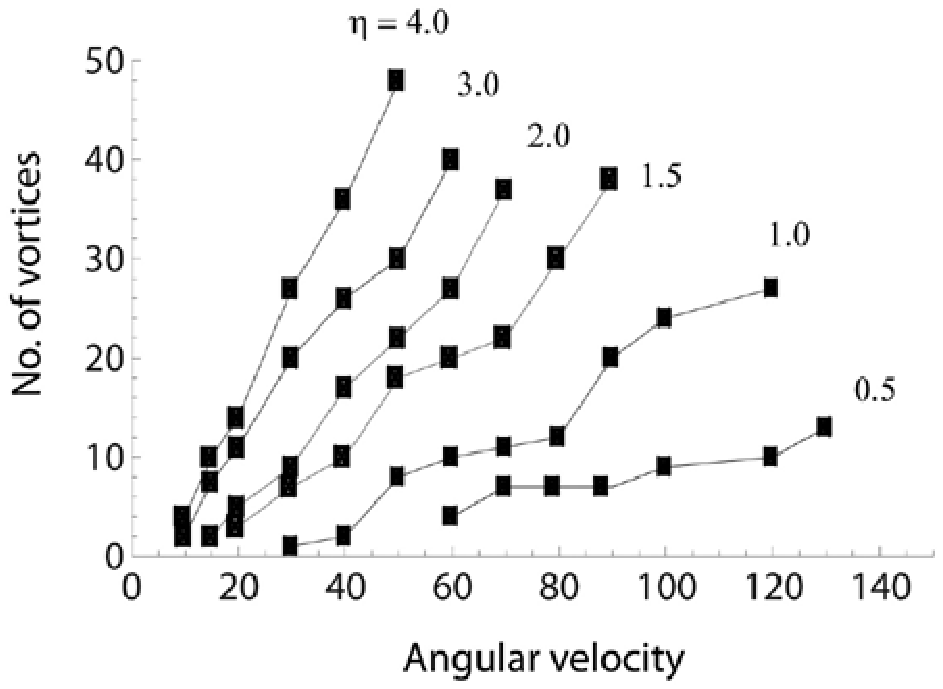}
\caption{Number of vortices around a galaxy, as a function of its angular
velocity, for different values of the coupling constant between galaxy and
scalar field.}
\end{center}
\end{figure}

\section{Discussions and outlook}

The basic idea of this model is that the universe is filled with a superfluid,
which is described by a vacuum complex scalar field, and dark matter is a
manifestation of density fluctuations of the superfluid. This idea has
similarities with previous theories that describe dark matter as a
Bose-Einstein condensate of some kind \cite{Harko, JWLee08, JWLee09, Popov1, Popov2,  HYLing1,  HYLing2, Suarez}. 
The difference is that our superfluid is relativistic and fills the entire universe, and, as described in
\cite{HLT1, HLT2}, had its origin in the big bang, and played a critical role in the
creation of matter through vortex activities. The big-bang theory gives a
Hubble parameter that decays in time like a power law. Such a behavior has
been postulated phenomenologically under the name of  ``intermediate
inflation", and compared with observations \cite{Barrow06}.

During the big-bang era, it was essential to use a scalar potential that is
asymptotically free, because the length scale was inflating rapidly. In the
present universe, however, we can describe the superfluid by a
phenomenological NLKG, with any convenient potential. The preliminary
investigation here aims mainly to show that it passes first encounters with
observations, and establish some parameters in the NLKG. Important limitations
of our model are the following: it treats the superfluid at absolute zero,
ignoring the normal fluid component; and it treats matter as external source,
ignoring its dynamics.

The normal fluid will consist of phonons of the Goldstone modes described
earlier, the quanta of the scalar field, with mass given by (\ref{quanta}),
and elementary particles coupled to the scalar field, for example the WIMPs
that particle physicists are searching for. If they exist, their excitation
from the superfluid would explain why they are expect to gather around
galaxies, but they are only secondary components of dark matter.

The present model cannot explain the velocity curve of Andromeda \cite{Rubin}, the
earliest indication of dark matter, which shows that ``dust" surrounding the
galaxy has a velocity too large to be accounted for by the visible galactic
mass. To do that, we need an equation of motion for matter.

In a more general sense, a extension of the present model should look towards
the inclusion of cosmic superfluidity into a hot big bang theory, and analysis
of the CMB. One would add the NLKG to any quantitative treatment. Clearly,
this would be major project yet to be undertaken.

We can only speculate on the effects of the cosmic superfluidity in the hot
big bang scenarios, and CMB fluctuations. They would have little influence on
small-scale phenomena, such as the nuclear and electromagnetic processes in
nucleosynthesis and barogenesis \cite{KolbTurner}, and the BAO (baryon acoustic
oscillations) in the CMB \cite{CMB2}. The main impact would be associated with
quantized vorticity, which could leave imprints on the CMB, and play a role in
structure formation in the universe.

\appendix

\section{Dimensionless form of NLKG}

For numerical computations, it is convenient to use the normalization
convention $\int d^{3}x\rho=1$, and absorb the mass of the source into $\eta$.
To put the equation of motion into dimensionless form, we scale all lengths
with the vacuum correlation length $\xi_{0}=\left(  \sqrt{\lambda}%
F_{0}\right)  ^{-1},$ and introduce dimensionless quantities indicated with an
overhead tilde:%
\begin{align}
\phi &  =F_{0}\tilde{\phi}\nonumber\\
t  &  =\xi_{0}\tilde{t}\nonumber\\
x  &  =\xi_{0}\tilde{x}\nonumber\\
\Omega &  =\xi_{0}^{-1}\tilde{\Omega}\nonumber\\
\rho &  =\xi_{0}^{-3}\tilde{\rho}\nonumber\\
\eta &  =\xi_{0}^{2}\tilde{\eta} \label{dimless}%
\end{align}
\newline The equation of motion for the scalar field, \ with an external
source $\rho$ representing the density of a galaxy (or star), which is
rotating with angular velocity $\Omega$, is given in dimensionless form by%
\begin{equation}
-\left(  1-2U\right)  \frac{\partial^{2}\tilde{\phi}}{\partial\tilde{t}^{2}%
}+\tilde{\nabla}^{2}\tilde{\phi}+\frac{\partial U}{\partial\tilde{t}}%
\frac{\partial\tilde{\phi}}{\partial\tilde{t}}+\tilde{\nabla}U\cdot
\tilde{\nabla}\tilde{\phi}-\left(  \left\vert \tilde{\phi}\right\vert
^{2}-1\right)  \tilde{\phi}-i\tilde{\eta}\tilde{\rho}\left(  \frac
{\partial\tilde{\phi}}{\partial\tilde{t}}+\mathbf{\tilde{\Omega}\times
\tilde{r}}\cdot\tilde{\nabla}\tilde{\phi}\right)  =0
\end{equation}
with the normalization convention%
\begin{equation}
\int d^{3}\tilde{x}\,\tilde{\rho}=1
\end{equation}
\newline The gravitational potential $U$ (which is dimensionless) can be
computed from Poisson's equation:%
\begin{align}
\nabla^{2}U  &  =4\pi G\left(  \rho_{\text{galaxy}}+\rho_{\text{scalar}}\right)
\nonumber\\
\rho_{\text{galaxy}}  &  =M_{\text{galaxy}}\rho\nonumber\\
G\rho_{\text{scalar}}  &  =\frac{G}{\lambda\xi_{0}^{4}}\left[  \left\vert
\frac{\partial\tilde{\phi}}{\partial\tilde{t}}\right\vert ^{2}+\left\vert
\tilde{\nabla}\tilde{\phi}\right\vert ^{2}+\frac{1}{2}\left(  \left\vert
\tilde{\phi}\right\vert ^{2}-1\right)  ^{2}\right]
\end{align}
The only adjustable parameter is the mass of the galaxy $M_{\text{galaxy}}$.
It is convenient to use the Milky Way as standard:%
\begin{equation}
GM_{\text{Milky Way}}=1.2\times10^{-15}\text{ cm}%
\end{equation}
Let%
\begin{equation}
\beta=\frac{M_{\text{galaxy}}}{M_{\text{Milky Way}}}%
\end{equation}
then the Poisson equation takes the form%
\begin{align}
\tilde{\nabla}^{2}U  &  =\beta c_{1}\tilde{\rho}+c_{2}\left[  \left\vert
\frac{\partial\tilde{\phi}}{\partial\tilde{t}}\right\vert ^{2}+\left\vert
\tilde{\nabla}\tilde{\phi}\right\vert ^{2}+\frac{1}{2}\left(  \left\vert
\tilde{\phi}\right\vert ^{2}-1\right)  ^{2}\right] \nonumber\\
c_{1}  &  =4.56\times10^{-40}\nonumber\\
c_{2}  &  =10^{-10}%
\end{align}


\end{document}